\documentclass[a4paper,preprint,preprintnumbers,amsmath,amssymb,superscriptaddress]{revtex4}

 \usepackage[T1]{fontenc}
 \usepackage{amsmath}
\usepackage{tabularx}
 \usepackage{graphicx}
\usepackage{multirow}
 \usepackage{SIunits}
 \usepackage{ulem}
 \usepackage{braket}
 \usepackage{color}
 \usepackage[draft]{fixme}
\usepackage{amssymb}
\usepackage{dcolumn}
\usepackage{bm}

\begin{document}

\title{Exciton fine-structure splitting of telecom wavelength single quantum dots: statistics and external strain tuning}

\author{Luca Sapienza \footnote{Present address: School of Physics and Astronomy, University of Southampton, Southampton SO17 1BJ, UK}}
\email{l.sapienza@soton.ac.uk}

 \author{Ralph N. E. Malein}
\author{Christopher E. Kuklewicz}
\author{Peter E. Kremer}
\affiliation{Institute of Photonics and Quantum Sciences, SUPA,
Heriot-Watt University, Edinburgh, United Kingdom}
 \author{Kartik Srinivasan}
\affiliation{Center for Nanoscale Science and Technology, National Institute of Standards and Technology, Gaithersburg, MD 20899, U.S.A.}
\author{Andrew Griffiths}
\author{Edmund Clarke}
\affiliation{EPSRC National Centre for III-V Technologies, University of Sheffield, United Kingdom}

\author{Ming Gong}
\affiliation{Department of Physics, The Chinese University of Hong Kong, Shatin, New Territories, Hong Kong, China}

\author{Richard J. Warburton}
\affiliation{Department of Physics, University of Basel, Klingelbergstrasse 82, CH-4056 Basel, Switzerland}

\author{Brian D. Gerardot}
 \email{b.d.gerardot@hw.ac.uk}
\affiliation{Institute of Photonics and Quantum Sciences, SUPA,
Heriot-Watt University, Edinburgh, United Kingdom}


\begin{abstract}
In a charge tunable device, we investigate the fine structure splitting of neutral excitons in single long-wavelength (1.1$\mu$m < $\lambda$ < 1.3 $\mu$m) InGaAs quantum dots as a function of external uniaxial strain. Nominal fine structure splittings between 16 and 136 $\mu$eV are measured and manipulated. We observe varied response of the splitting to the external strain, including positive and negative tuning slopes, different tuning ranges, and linear and parabolic dependencies, indicating that these physical parameters depend strongly on the unique microscopic structure of the individual quantum dot. To better understand the experimental results, we apply a phenomenological model describing the exciton polarization and fine-structure splitting under uniaxial strain. The model predicts that, with an increased experimental strain tuning range, the fine-structure can be effectively canceled for select telecom wavelength dots using uniaxial strain. These results are promising for the generation of on-demand entangled photon pairs at telecom wavelengths.
\end{abstract}

 \maketitle

Remarkable progress in the field of self-assembled quantum dots (QDs) has been made in the last decade, primarily using InGaAs QDs emitting at $\lambda$ < 1 $\mu$m. One such noteworthy result is the demonstration of on-demand polarization entangled photons from a single QD via the biexciton-to-exciton-to-vacuum state cascade \cite{Bennett, Pascale, Gershoni, Shields}. However, deterministic photon sources at telecom wavelengths \cite{singleph_tel, tel_FSS, PhC, DWELL, low_counts, telecom_InP, Tarta} are required for efficient communication via fiber, free-space through the atmosphere, or for integration with silicon photonics. Unfortunately, due to materials challenges for long-wavelength QDs and traditional difficulties in photon detection at telecom wavelengths, to date relatively little progress has been made with QDs in the telecom O-band ($\lambda$  $\sim$ 1310 nm) or C-band ($\lambda$  $\sim$ 1550 nm) compared to shorter wavelength ($\lambda$ < 1 $\mu$m) QDs. Different approaches have been used in order to overcome the challenge of telecom photon detection. For instance, photonic crystal cavities \cite{PhC} or fiber taper waveguides \cite{DWELL} have been used to enhance the spontaneous emission rate and channel the emitted light into a specific optical mode. Frequency up-conversion from telecom to visible wavelength has also been implemented \cite{Rakher} and single-photon superconducting detectors are being developed to achieve a more efficient detection at telecom wavelengths \cite{Mike}.

In typical self-assembled QDs, the lattice symmetry is broken from $T_d$ to $C_{2v}$ due to macroscopic structure anisotropy, and from $C_{2v}$ to $C_1$ symmetry due to other nonuniform effects, including local strain, alloys, interface effects, etc. For quantum dots with $C_{2v}$ or $C_1$ symmetry,  two bright exciton states belonging to different irreducible representations arise due to electron-hole exchange interaction \cite{Bester1, Bester, Gong, Luo}. This non-degenerate doublet  is referred to as a fine-structure splitting (FSS) (see Fig. 2a).
The FSS doublet is orthogonally polarized in the linear basis and leads to distinguishability in the biexciton-to-exciton-to-vacuum cascade. The magnitude of the FSS is determined by anisotropy in the strain, shape and composition of the dot, as well as from the crystal inversion asymmetry \cite{Bester1, Bester, Gong}. If the FSS is smaller than the homogeneously broadened emission linewidth \cite{Shields}, the biexciton-to-exciton-to-vacuum cascade can lead to the emission of polarization-entangled photon-pairs \cite{Bennett, Pascale, Gershoni}.
FSS of the order of a few tens of $\mu$eV have recently been manipulated and/or canceled in QDs emitting at $\lambda <$ 1  $\mu$m via an electric field \cite{Bennett, lateral_E, Kowalik}, uniaxial strain \cite{strain1, strain2, Chris} or combined electric field and strain \cite{Trotta}.

Here we characterize the FSS in long-wavelength QDs and investigate the prospect of canceling it using uniaxial strain. We investigate two samples containing QDs emitting photons near the telecom O-band: one sample consists of QDs in the bulk and the other is a charge tunable QD device \cite{Warburton}. In the second sample, deterministic charging allows clear identification of the charged excitonic states visible in the photoluminescence (PL) spectra which allows us to selectively address single exciton and biexciton lines. By carrying out polarization-resolved PL we measure FSS as low as a few tens of $\mu$eV. By applying uniaxial strain \cite{strain1, Chris, strain2, Trotta}, we demonstrate manipulation of the FSS and reveal different critical stresses, $p_c$ (defined as the stress required to reach the minimal FSS), and minimal FSS for different QDs. Tantalizingly, application of the empirical model of Ref. \cite{Gong} predicts that the effective cancellation of FSS using uniaxial strain is achievable for select QDs investigated here. For the remaining dots characterized here, the incorporation of a second tuning knob \cite{Trotta}, e.g. electric fields \cite{Bennett,  lateral_E, Kowalik, Trotta} or another strain axis \cite{extra_strain}, should enable the realization of a source of polarization entangled photons at telecom wavelengths.

The samples consist of a single layer of self-assembled InAs QDs in an In$_{0.18}$Ga$_{0.82}$As quantum well (dot-in-a-well, or DWELL structures). We have characterized our sample using transmission electron microscopy (TEM) and a typical image is shown in Fig.1c. The QDs were grown within a quantum well in order to promote the relaxation of the structure during the growth and achieve larger sizes (and therefore longer emission wavelength) than typical near infrared InAs/GaAs QDs. This is confirmed by the TEM picture from which we can estimate a lateral size of about 18-25 nm and a height of about 8-13 nm  for single QDs. We note that such larger values in the height compared to shorter wavelength QDs are expected to enable larger tunability in the FSS under applied vertical electric field \cite{Luo}. The DWELL structure red-shifts the emission wavelength of the QDs to a wavelength range between 1080 and 1310 nm at T = 4 K \cite{DWELL, low_counts, Tarta}. As these QDs have deeper confinement potentials than QDs at  $\lambda$ < 1 $\mu$m, a reduced tunnel barrier thickness (14 nm) is required to obtain sharp charge state transitions in PL characterization as a function of applied bias in the charge tunable device (see Fig. 1a) \cite{Ediger}. The structure, shown in Fig. 1a, has a relatively large (104 nm) capping layer separating the QD and the AlGaAs superlattice to minimize the effect of localized defects at the AlGaAs interface \cite{Julien}. This device geometry gives a lever arm, defined as the ratio between the device length (400 nm) and the tunnel barrier thickness (15 nm), of $\sim$ 27 which results in an operating voltage of $\sim$ -7.5 V for charging the QD ground state.

We optically excite the single QDs by using a non-resonant continuous-wave laser ($\lambda$ = 830 nm) and collect the emitted photons with confocal micro-PL. A zirconia super-solid immersion lens (SIL) is positioned on the surface of the sample to increase  the collection efficiency and reduce the excitation and collection spot size \cite{superSIL}. With the super SIL, we obtain saturation counts up to $\sim$ 300 Hz on a liquid-nitrogen cooled InGaAs detector array (equivalent to a photon count rate of $\sim 2\times 10^{4}$ Hz). Our spectrometer has a resolution of 0.10 nm (75 $\mu$eV) at 1300 nm and, using a double Lorenztian fit to the emission lines, we are able to resolve the FSS with a few $\mu$eV resolution. The samples under study have a high QD density and spectrally isolated QDs can be found at the tails of the size distribution (between $\sim$ 1080-1130 and $\sim$ 1240-1310 nm at T = 4 K).
To apply the uniaxial strain (along the [110] crystallographic axis), we glue the sample to a piezoelectric lead zirconia titanate (PZT) ceramic stack to which a bias (V$_{PZT}$) from -300 to +300 V can be applied. These voltages correspond to an upper bound for the applied strain of $\sim\pm$ 13.9 MPa (for details on the strain calibration, see \cite{Chris}).

An example of the PL spectra as a function of the voltage applied to the sample (V$_{gate}$) is shown in Fig. 1b. Discrete jumps of the emission lines are clearly visible, a signature of Coulomb blockade \cite{Warburton}. The lineshapes of the emitted spectra reveal resolution limited linewidths of about 0.08 nm and confirm the high optical quality of the samples. To unambiguously identify the neutral exciton (X$^0$), biexciton (2X$^0$) and single negatively charged exciton (X$^{1-}$) emission lines, we perform polarization dependent PL. An example of the spectra for orthogonal polarizations are shown in Fig. 2a: the emission lines at $\sim$ 1285.3 and $\sim$ 1286.5 nm show FSS, while the line at $\sim$ 1288.9 nm does not shift with changing polarization. For orthogonal polarizations one peak shifts towards shorter and one towards longer wavelengths as expected for X$^0$ ($\lambda$ $\sim$ 1285.3 nm) and 2X$^0$ ($\lambda$ $\sim$ 1286.5 nm). Due to the Coulomb blockade signature (see Fig. 1b) and the absence of any FSS, the emission line at $\sim$ 1288.9 nm is attributed to the X$^{1-}$ recombination from the same QD.

Combining the statistics of the measured FSS from both samples, we see a full range of FSS between 16 and 136 $\mu$eV for 76 measured QDs (Fig. 2b). This range of FSS is considerably smaller than previous reports on FSS for QDs emitting at similar wavelengths \cite{large_FSS, tel_FSS, Tarta}, an important result as a smaller initial FSS requires more modest external fields for complete cancellation. We do not observe a clear correlation between the emission wavelength and the FSS as has been observed for both strained \cite{Seguin} and unstrained \cite{Abbarchi1, Abbarchi2, Plumhof} QDs at shorter emission wavelengths. For unstrained dots, increasing FSS was observed as the QD size increased and generally attributed to dot morphology as larger dots have increased shape anisotropy. One signature of strong shape anisotropy is preferential alignment of the polarization axes of the FSS with a crystallographic direction. Therefore, in Fig. 2c we present the polarization angles of the high and low-energy FSS peaks for the dots we measured. We observe that QDs at all wavelengths in the charge tunable device and QDs at shorter wavelengths in the bulk sample tend to align along the crystallographic axes, whereas longer wavelength QDs in the bulk sample display more random FSS polarization orientations.

\begin{table}
\caption{Strain tuning of single QDs. The wavelength $\lambda_{V_{PZT=0}}$ is the central wavelength of the excitonic line without applied external strain. The FSS slope is the result of a linear fit of the FSS splitting in the full V$_{PZT}$ range, except for QDs 2, 9, 12 where only the points in the linear regime were fitted (see Fig. 3). The angle $\theta$ represents the polarization angle of the low energy peak with respect to the [110] crystallographic axis. FSS$_{min}^{exp}$ is the minimal value of FSS that we measure in our experiments. $\Delta E$ refers to the energy shift for the full tuning range for increasing tensile strain.  $2|\delta|$ and $2|\kappa|$ refer to the diagonal and off-diagonal lower bounds for FSS, respectively. (Note that 1 V$_{PZT}$ = 46 KPa.)}
\centering 
\setlength{\tabcolsep}{3pt}
\begin{tabularx}{\textwidth}{c c c c c c c c c c} 

\hline\hline 
 QD & $\lambda_{V_{PZT}=0}$  & FSS$_{V_{PZT}=0}$  & FSS slope  & $\Delta$ FSS  & FSS$_{min}^{exp}$& $\Delta E$ & $\theta_{V_{PZT}=0}$ &  $2|\delta|$ & $2|\kappa|$\\  ($\#$)&(nm)&($\mu$eV)&($\mu$eV/V$_{PZT}$)&($\mu$eV)&($\mu$eV) &(meV)&($\degree$)&($\mu$eV)&($\mu$eV) \\ [1ex] 

\hline 
1 & 1167.0 & 45.2$\pm$2.1 & 0.015 & 8.3&  35.7$\pm$0.8 &0.91&83.6&44.1$\pm$2.3&10.0$\pm$0.5\\
2 & 1224.7 & 23.5$\pm$2.4 & -0.074 & 15.1 & 20.1$\pm$1.1   & 0.99&-5.0&23.1$\pm$2.3&4.1$\pm$0.5\\
3 & 1227.3 & 41.0$\pm$3.4& -0.036 & 21.8 & 23.1$\pm$1.4 & 0.82&-2.3&40.9$\pm$3.4&3.3$\pm$0.3\\
4 & 1228.0 & 46.0$\pm$5.9 & -0.022 & 12.4& 37.5$\pm$2.6&0.82&-2.9&45.8$\pm$5.9&4.6$\pm$0.6\\
5& 1234.0 & 39.5$\pm$1.1 & 0.026 &15.6& 28.9$\pm$0.6&0.81&-0.1&39.5$\pm$1.1&0.1$\pm$0.0\\
6 & 1234.4 & 34.4$\pm$0.9 & 0.024 & 13.0&29.3$\pm$0.6&0.81&0.8&34.4$\pm$0.9&1.0$\pm$0.0\\
7 & 1235.3 & 32.2$\pm$0.7 & 0.022 & 13.8&21.5$\pm$0.6&0.81&-1.5&32.2$\pm$4.7&1.7 $\pm$0.0\\
8& 1241.2 & 47.2$\pm$2.1 & 0.077 & 46.4&22.5$\pm$0.4&0.72&-4.7&46.6$\pm$2.1&7.7$\pm$0.1\\
9 & 1267.0 & 49.7$\pm$5.1 & 0.051 & 19.9 & 49.0$\pm$2.6 & 0.70&1.3&49.6$\pm$5.1&2.3$\pm$0.2\\ 
10& 1285.7 & 50.3$\pm$2.6 & -0.021 & 14.2&47.3$\pm$1.4& 0.75&75.2&43.7$\pm$2.6&24.8$\pm$0.3\\
11& 1288.0 & 47.2$\pm$1.2 & -0.017 & 10.3&39.1$\pm$1.0&0.90&-8.6&45.1$\pm$1.1&14.0$\pm$0.4\\
12& 1296.2& 68.6$\pm$2.5 & -0.024 & 11.1&63.3$\pm$1.7&0.74&-9.9&64.5$\pm$2.4
&23.2$\pm$0.8\\[1ex] 

\hline 

\label{strain_table} 

\end{tabularx}

\end{table}

We next apply an external uniaxial strain and find that the FSS can be manipulated in a reversible way and that significant reductions of the FSS can be achieved (see Fig. 3).  Table \ref{strain_table} summarizes the results from 12 single QDs. We observe tuning ranges ($\Delta$ FSS) from 8.3 to 46.4 $\mu$eV, slopes ranging from -0.074 to 0.077 $\mu$eV/V$_{PZT}$, and blueshifts of the emission energy $\Delta E$ of $\sim$ 1 meV for increasing tensile strain. In Fig. 3a, we show a polar plot for the two orthogonally polarized exciton lines for QD10. As shown, going from -300 to +300 $V_{PZT}$, the alignment of the polarization angle $\theta$ with respect to the [110] axis only varies by a few degrees, a typical result in our experiments. While most of the QDs under study show a linear dependence of the FSS as a function of the applied strain (see Fig. 3b left panel), for QD2 we observe a parabolic modification of the FSS which reaches a minimum (FSS$_{min}^{exp}$) of 22.4 $\pm$ 2.2 $\mu$eV (see Fig. 3b right panel).

The application of uniaxial strain is expected to modify the FSS in a quadratic way, with the minimum of the parabola representing the minimal FSS that is reachable for a specific QD \cite{Bennett, Trotta}. The critical stress required to reach the minimum FSS depends on the shape and composition of each specific QD \cite{Bester, Gong}. If $p_c$ is not experimentally reachable, one observes a linear response with either positive or negative tuning slopes, depending on which arm of the parabola is probed (see Fig. 3b). The realization of a larger strain range would enable the minimum of the parabola to be reached for each dot.

The FSS is a result of the asymmetric confining potential of the carriers trapped within the quantum dot. This symmetry lowering can be attributed to different factors: shape anisotropy, the presence of piezoelectric fields (due to strain from the different lattice constants of the materials composing the QD structure that separates negative and positive charge centers), and different interface potentials (due to differences in the interfaces at the atomistic level). This last effect is related to the position of the atoms in the nanostructure and, therefore, is the most sensitive to applied strain. As shown in Refs. \cite{Bester} and \cite{Gong}, the application of external strain does not change the macroscopic shape of the quantum dot considerably (less than 0.2\%). Also, piezoelectricity seems to have a marginal effect in the theoretical evaluation of the FSS under strain \cite{Gong}. Hence, we conclude that the experimental results revealing very different dependencies for the FSS on the applied strain for each dot are caused by uniqueness at the atomistic level. The fact that FSS$_{min}^{exp}$ agrees with the results of the model of Ref. \cite{Gong} (see Table \ref{strain_table} and discussion below) further supports these conclusions.

The behavior of the FSS under uni-axial strain can be understood using the basic picture presented in Ref. \cite{Gong}. Using the same notation, the effective bright exciton Hamiltonian reads $H = (\delta + \alpha p/2) \sigma_z  + (\kappa + \beta p) \sigma_x$, where $p$ is the external stress, $\alpha$, $\beta$, $\kappa$ and $\delta$ are empirical parameters that depend strongly on the microscopic structure of the QDs, and $\sigma_{x}$, $\sigma_z$ are the Pauli matrices. The FSS then reads as
\begin{equation}
\Delta = \sqrt{4(\beta p + \kappa)^2 + (\alpha p + 2\delta)^2}
\end{equation}
Generally, for stress along either [110] or [1$\bar{1}$0] direction, $\alpha \ne 0$ and $\beta = 0$ (see Table 1 in Ref. \cite{Gong}) and the lower bound of FSS can be reached
when the diagonal elements are removed, i.e., $\delta + \alpha p/2 = 0$. We call this lower bound $\Delta_{min} = 2|\kappa|$ the "off-diagonal lower bound". For stress along [100] or [010] direction, the lower bound of FSS can be reached when the off-diagonal elements are removed, $\kappa + \beta p = 0$, and we call this lower bound $\Delta_{min} = 2|\delta|$ the "diagonal lower bound". The lower bound of FSS can thus be predicted using the FSS $\Delta$ (labeled FSS$_{V_{PZT}=0}$ in Table \ref{strain_table}) and polarization angle $\theta$ at zero bias using
\begin{equation}
\delta = \Delta \cos(2\theta)/2, \quad \kappa = - \Delta \sin(2\theta)/2
\end{equation}
Here we compare our results to this phenomenological model. FSS$^{exp}_{min}$ and the predicted diagonal ($2|\kappa|$) and off-diagonal ($2|\delta|$) lower bounds are presented in Table \ref{strain_table}. Note that the minimum of the parabola is reached for QD2 only, therefore the other values reported do not represent the minimal achievable FSS for the QDs under study, but the minimal FSS achieved under the current experimental conditions. In general, we find FSS$^{exp}_{min} > 2|\kappa|$ as expected due to $|p_c|$ exceeding maximum range of the experimentally applied stress. Additionally, there might be a non-uniformity of the external stress in the experiment that results in the applied strain not exactly oriented along the [110] or [1$\bar{1}$0]. In this scenario, $\alpha \ne 0$ and $\beta \ne 0$ and one expects $2|\delta|$ > FSS$^{exp}_{min}$ > $2|\kappa|$. Notably, applying additional stress along [100] components can further reduce the FSS and the application of two independent external stress is expected to cancel the FSS \cite{extra_strain}.

QDs with $\theta$ aligned along the [110] or [100] directions are expected to reach the smallest FSS when an external stress is applied \cite{Gong}. In contrast to shorter wavelength ($\sim$ 950 nm) smaller QDs whose alignment is more random \cite{Seidl_E}, we observe that the long wavelength charge-tunable QDs measured here are well aligned with the [110] axis (see Fig. 2c and Table \ref{strain_table}). No post selection has been done to select QDs better aligned to the crystallographic axis. In fact, for QDs 5 and 6 in Table \ref{strain_table}, $2|\kappa| \le 1 \micro$eV, the typical transform limited linewidth for self-assembled QDs. This is significant: with a larger strain tuning range, entangled photon pair generation at telecom wavelength should be possible. For QDs in which $2|\kappa| >$ 1 $\micro$eV,  a second external field will allow complete cancellation of the FSS \cite{Trotta, extra_strain}. The small rotations of $\theta$ shown in Fig. 3a are expected when the FSS varies linearly with the applied strain. We also note that for QD2 the rotation of $\theta$ is still limited to $\sim$ 5$\degree$, even though the minimal FSS is reached. Polarization rotations smaller than the ones reported in References \cite{Bennett} and \cite{Trotta} have been predicted for QDs with different shapes and composition \cite{Gong, Bester}. One possible explanation for the experimental observation of limited $\theta$ rotation for QD2 is that the deep confinement potential of the telecom wavelength QDs reduces penetration of the carrier wavefunctions into the barrier material, leading to reduced sensitivity to the QD environment (e.g. alloy disorder at the interface) \cite{Mlinar, Luo} and, therefore, less pronounced rotations of $\theta$. Further investigation, which goes beyond the scope of our current work, is required to correlate the dot's morphology with the FSS and $\theta$. One promising approach based on the statistical trends of an ensemble of dots has recently been developed and applied to shorter wavelength QDs \cite{Gong2}.

In conclusion, we have realized a charge tunable structure for QDs emitting at telecom wavelengths to enable deterministic charging of the neutral exciton. By performing polarization-resolved PL, we observe nominal FSS of neutral exciton lines down to 16 $\mu$eV. We demonstrate that the application of uniaxial strain allows significant manipulation of the FSS and we observe linear reductions of the FSS for most of the QDs. Each QD shows a unique response to the applied strain, which is attributed to different structural properties of the QDs that result in different values of $p_c$. Further, we have applied an empirical model to describe the polarization and FSS under uniaxial strain, which predicts that the FSS can be effectively canceled for some QDs investigated here, thus enabling deterministic entangled photon pair generation at telecom wavelengths. These results are a promising step in bridging the gap in the state-of-the-art between mature QDs emitting at $\lambda < 1 \mu$m and telecom wavelength QDs.

The authors would like to thank A. Dada for comments on the manuscript and acknowledge the financial support for this work from the Royal Society, EPSRC, the ERC, and NCCR QSIT. M.G. is supported in part by Hong Kong RGC/GRF (401512, 401113) and Hong Kong Scholars (XJ2011027).

\newpage

\begin{figure}[tb]
   \centering
  \includegraphics[height=17cm, width=0.8\linewidth]{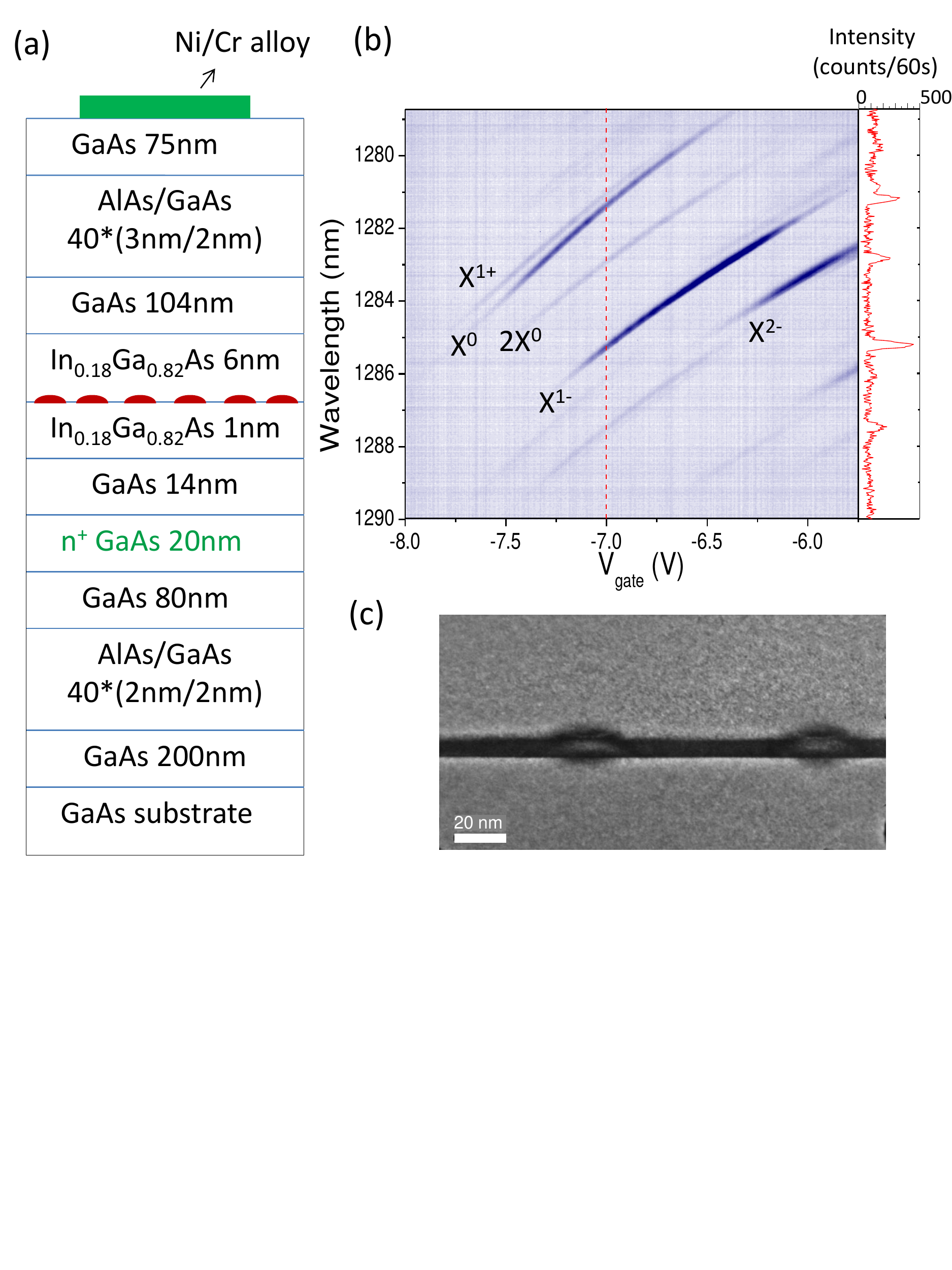}
   \caption{(a) Schematic of the charge tunable structure. The red symbols represent the QD layer. (b) PL spectra collected as a function of the applied gate voltage under non-resonant excitation ($\lambda$ = 830 nm) at a temperature T = 4 K. The peaks corresponding to the emission from different states of a single QD are labeled accordingly. The right panel shows a linecut of the contour plot at the corresponding red dashed line, at V$_{gate}$ = 7 V. (c) Transmission electron microscopy image of two quantum dots grown in the charge tunable device (image provided by Richard Beanland, Integrity Scientific Ltd).}

   \label{fig1}
\end{figure}

\begin{figure}[tb]
   \centering
  \includegraphics[width=0.75\linewidth]{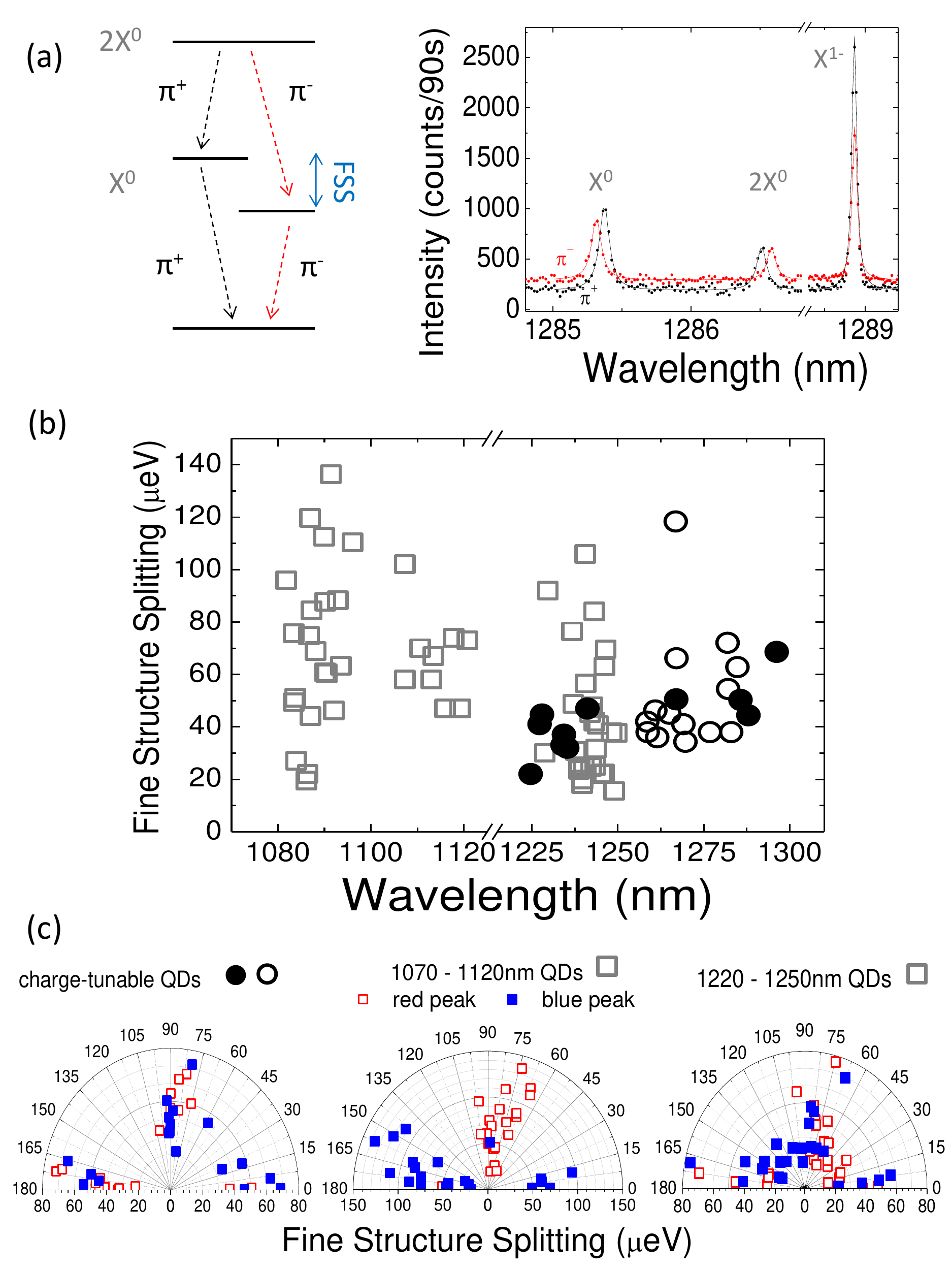}
   \caption{(a) Left panel: schematic of the biexciton (2X$\textsuperscript{0}$)-to-exciton (X$\textsuperscript{0}$)-to-vacuum state transitions and their respective polarisations ($\pi\textsuperscript{+}$ or $\pi\textsuperscript{-}$). Right panel: An example of PL spectra at orthogonal polarisations, showing the X$\textsuperscript{0}$, 2X$\textsuperscript{0}$ and singly charged exciton (X$\textsuperscript{1-}$) emission lines (full symbols) and a Lorentzian fit to the data (solid lines). (b) FSS measured on single QDs with no applied external strain. The circles represent values obtained from the charge tunable device (the full symbol correspond to the 12 QDs in Table I), while the open squares correspond to measurements from a layer of DWELL QDs grown in the bulk. (c) Polarisation angle of the short (blue full symbol) and long (red open symbol) wavelength peak with respect to the [110] crystallographic axis for the exciton-to-vacuum transition.}
   \label{fig2}
\end{figure}

\begin{figure}[tb]
   \centering
  \includegraphics[height=17cm, width=0.8\linewidth]{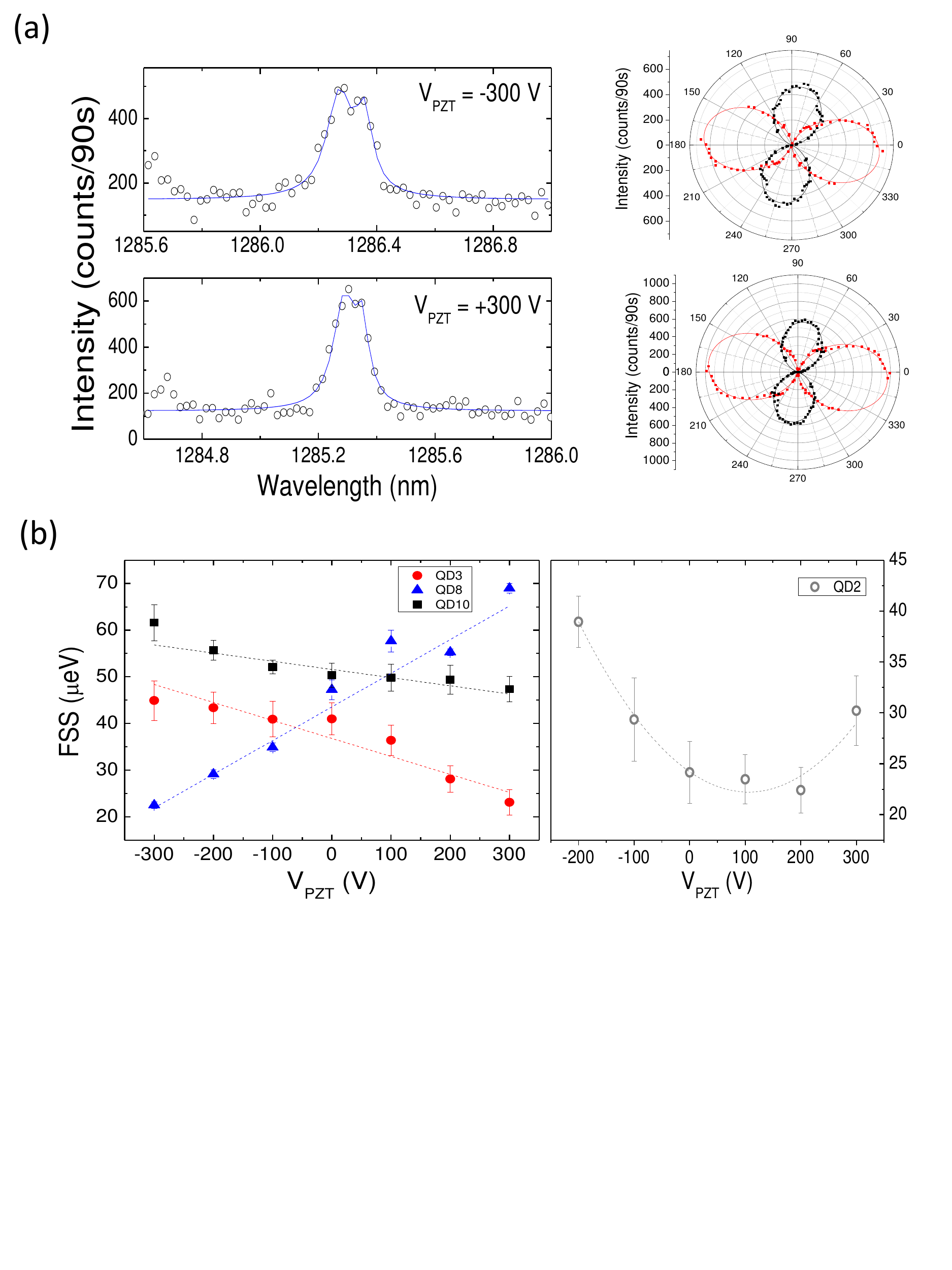}
   \caption{(a) Left panel: Example results of the manipulation of the X$^0$ FSS (from 61.6 to 47.3 $\mu$eV) of QD10 with uniaxial strain. The solid lines are double Lorentzian fits to the data, collected at a polarisation   angle of $\sim$45$\degree$ (open circles). Right panel: Two examples of polar plots for the two orthogonal exciton lines at V$_{PZT}$ = -300, +300 V. The solid lines are fits to the data. (b) FSS as a function of applied voltage on the PZT stack for four different single QDs. The error bars are the standard deviation from the mean value of the FSS, obtained from 43 fits to the experimental spectra collected as a function of polarisation angle ranging between 0$^{\circ}$ and 140$^{\circ}$. The dashed lines in the left (right) panel are linear (quadratic) fits to the data.}
\label{fig3}
\end{figure}

 \end{document}